\documentstyle[11pt,aaspp4,psfig]{article}

\def\arcsec              {$^{\prime\prime}$}

\def\Dcal{{\cal D}}
\def\Fcal{{\cal F}}
\def\Tcal{{\cal T}}
\def\Scal{{\cal S}}
\def\Rcal{{\cal R}}
\def\Kcal{{\cal K}}
\def\Hcal{{\cal H}}
\def\Mcal{{\cal M}}

\def\Ccal{{\cal C}}
\def\Lcal{{\cal L}}
\def\Wcal{{\cal W}}

\begin{document}

\title{A method for spatial deconvolution of spectra}

\author{F. Courbin}

\affil{
Universidad Cat\'olica de Chile,\\
Departamento de Astronomia y Astrofisica,
Casilla 306, Santiago 22, Chile;\\
Institut d'Astrophysique et de G\'eophysique de Li\`ege,  
Avenue de Cointe 5, B-4000 Li\`ege, Belgium;\\
fcourbin@astro.puc.cl}

\vspace*{5mm}

\author{P. Magain\altaffilmark{1}, M. Kirkove\altaffilmark{2}, S. Sohy}

\affil{Institut d'Astrophysique et de G\'eophysique de Li\`ege, Avenue
de  Cointe   5,  B-4000 Li\`ege,  Belgium;\\  Pierre.Magain@ulg.ac.be,
Murielle.Kirkove@ulg.ac.be, Sandrine.Sohy@ulg.ac.be}

\altaffiltext{1}{Also Ma\^{\i}tre de Recherches au FNRS (Belgium)}
\altaffiltext{2}{Aspirante au FNRS (Belgium)}

\begin{abstract}

A  method   for spatial  deconvolution of spectra    is presented.  It
follows    the same fundamental    principles    as the  ``MCS   image
deconvolution  algorithm''  (Magain,  Courbin, Sohy,   1998) and  uses
information  contained  in the spectrum of   a  reference Point Spread
Function (PSF)  to   spatially deconvolve  spectra  of  very   blended
sources.  An improved resolution rather than  an infinite one is aimed
at, overcoming the well  known problem of ``deconvolution artefacts''.
As in  the MCS  algorithm,  the data  are   decomposed into a  sum  of
analytical point sources  and a numerically deconvolved background, so
that the  spectrum of  extended  sources in the immediate  vicinity of
bright  point sources may be accurately  extracted and sharpened.  The
algorithm has been tested on simulated data including seeing variation
as a  function of wavelength and  atmospheric refraction.  It is shown
that  the spectra of  severely  blended point sources  can be resolved
while fully preserving  the spectrophotometric properties of the data.
Extended   objects ``hidden'' by  bright   point  sources (up to   4-5
magnitudes brighter) can be accurately recovered as well, provided the
data have a sufficiently high total signal-to-noise ratio (200-300 per
spectral resolution element).   Such spectra  are  relatively easy  to
obtain,   even  down  to  faint magnitudes,    within a  few  hours of
integration time with 10m class telescopes.

\end{abstract}


\clearpage

\section{Background and motivation}

Tremendous efforts  have been devoted to  the development of numerical
methods aimed at improving  the   spatial resolution of   astronomical
images.  However, the  techniques proposed  so  far and  most commonly
used (e.g., Richardson, 1972, Lucy,  1974,  Skilling \& Bryan,  1984),
tend    to   produce    the    so-called  ``deconvolution  artefacts''
(oscillations in the vicinity  of  high spatial frequency  structures)
and    to  alter the photometric  and    astrometric properties of the
original  data.  Recently, Magain,  Courbin  \& Sohy (1998;  hereafter
MCS) proposed and   implemented  a new deconvolution  algorithm  which
overcomes such  drawbacks.  Its success is mainly the consequence of a
deliberate choice to achieve  an {\sl improved} resolution rather than
an   infinite  one, hence   avoiding  to  retrieve spatial frequencies
forbidden by the  sampling  theorem.  Many successful  applications of
the algorithm have been carried  out in the  framework of an intensive
effort  to obtain  detailed   light/mass maps for  lensing  galaxies
(e.g.,  Courbin  et al.  1997,  1998, Burud  et  al.  1998a,b).  These
results, when  compared with more recent  Hubble Space Telescope (HST)
images, demonstrate the efficiency of  the method to produce  reliable
high resolution images.

Undoubtedly,  high spatial resolution plays a  key role in most of the
major  advances made in observational astrophysics.   This is true not
only in imaging, but also in spectroscopy.  We present in this paper a
spectroscopic   version of the  MCS  algorithm and we demonstrate from
realistic simulations that flux calibrated spectra of severely blended
point   sources can be accurately  recovered.   We  also  show how the
algorithm can   be used  to  decontaminate  the spectrum   of extended
objects from light pollution by  very nearby, eventually bright, point
sources.

\section{Spatial deconvolution of spectra}


 MCS have shown that sampled images should not be deconvolved with the
observed Point Spread Function  (PSF),  but with a  narrower function,
chosen so that the final  deconvolved image  can be properly  sampled,
whatever sampling step is adopted to represent  it.  For this purpose,
one can {\sl choose} the final PSF $\Rcal(\vec{x})$ of the deconvolved
image and {\sl compute} the PSF $\Scal(\vec{x})$  which should be used
to   perform   the deconvolution instead  of   the  {\sl observed} PSF
$\Tcal(\vec{x})$.   The profile  $\Scal(\vec{x})$  may be obtained  by
inverting the equation (see Sect. 3.):

\begin{eqnarray}
\Tcal(\vec{x}) = \Scal(\vec{x}) \ast \Rcal(\vec{x}),
\label{equation:tsr}
\end{eqnarray}

 where the  symbol  ``$\ast$'' stands for  the  convolution.  Equation
(\ref{equation:tsr}) should   always be considered together   with the
important constraint that all three profiles must be properly sampled.

A straightforward  consequence of {\sl choosing}  the shape of the PSF
$\Rcal(\vec{x})$ is   that it is,  indeed,   exactly known.  Such {\sl
prior knowledge} can be used to decompose the data (image or spectrum)
into a sum of point sources with known analytical spatial profile, and
a  deconvolved    numerical  background.    This   decomposition   was
successfully used in the  MCS image deconvolution algorithm.  We apply
the same fundamental  rule to construct  an algorithm for  the spatial
deconvolution of spectra.

\subsection{Constructing the algorithm}

 A 2-D spectrum  can be described  as $M$ spectral resolution elements
(e.g., $M$ lines), composed of  $N$ spatial resolution elements (e.g.,
$N$ columns).  Each spectral resolution element of the spectrum can be
approximated as a quasi-monochromatic 1-D image (see Appendix) that we
decompose as  in  the MCS image   deconvolution algorithm.  The  pixel
intensities $\Fcal(\vec{x})$ of such a 1-D image may then be written as

\begin{eqnarray}
\Fcal(\vec{x}) = \Hcal(\vec{x}) + \sum_{k=1}^{N_{\star}} a_k 
\Rcal(\vec{x}-\vec{c_k}),
\label{equation:decomp}
\end{eqnarray}

which     is the sum   of  a    1-D numerical deconvolved   background
$\Hcal(\vec{x})$   and  of $N_{\star}$  profiles $\Rcal(\vec{x})$ with
intensities $a_k$ and  centers $c_k$.  The profile $\Rcal(\vec{x})$ is
chosen  to be Gaussian, with  fixed width (i.e., resolution) all along
the spectral direction.   The final deconvolved spectrum  is therefore
corrected for    seeing variations   with wavelength.   Moreover,  the
spectra may suffer from slit misalignment with respect to the physical
dimensions  of the  detector and  from atmospheric  refraction.  As  a
consequence, the position  of a given point source  on the detector is
wavelength dependent.   The deconvolved  (and decomposed) 2-D spectrum
which matches   the data  at   best, may  therefore  be   obtained  by
minimizing the function

\begin{eqnarray}
\Ccal_{\chi^2} & = &\sum_{j=1}^{M} \sum_{i=1}^{N} \frac{1}{\sigma_{i,j}^2} 
\left[ \Scal_{j}(\vec{x})\ast 
\left( \Hcal_{j}(\vec{x}) + \sum_{k=1}^{N_{\ast}} a_{k,j} \, \Rcal(\vec{x} 
\! - \! \vec{c}_{k,j})
\right) - \Dcal_{j}(\vec{x}) \right ]_{i}^2,
\label{equation:MCSspec}
\end{eqnarray}

where   $\Dcal_{j}(\vec{x})$ is the  $j^{th}$  spectral element of the
data spectrum,  $\Scal_{j}(\vec{x})$ is  the $j^{th}$ spectral element
of the (narrower)    PSF.   Note that  the    width of the     profile
$\Scal(\vec{x})$ may  vary   with  wavelength, as  does   the observed
profile $\Tcal(\vec{x})$,  in  order to ensure  that  $\Rcal(\vec{x})$
does not.  The sum over  $i$ is the summation  of the $N$ pixel values
along the spatial direction of the $M$ spectral resolution elements.


Importantly, the  deconvolution  has  to    be  performed under    the
constraint that the  deconvolved background $\Hcal(\vec{x})$ is smooth
on the  length  scale  of  the final   resolution (represented by  the
profile $\Rcal(\vec{x})$).  This is efficiently done by minimizing

\begin{eqnarray}
\Lcal_{1} & = &\sum_{j=1}^{M} \sum_{i=1}^{N} \left [ 
\Hcal_{j}(\vec{x}) - \Rcal(\vec{x})\ast \Hcal_{j}(\vec{x}) 
\right ]_{i}^2
\label{equation:smooth}
\end{eqnarray}

where  the   notations  and indices   are   the same  as  in  equation
(\ref{equation:MCSspec}).    This smoothing  applies  in   the spatial
direction only and independently for each spectral resolution element.

In   spectroscopy, one may take    advantage  of an additional   prior
knowledge  available: the fact  that  the  position of  a given  point
source at a given wavelength is highly correlated with its position in
the neighbouring  spectral  resolution  elements.  We  introduce  this
prior knowledge as a second constraint:

\begin{eqnarray}
\Lcal_{2} & = &\sum_{j=1}^{M} \left [ 
c_{j} - \sum_{j'=-W/2}^{W/2} g_{j'}c_{j+j'}
\right ]^2
\label{equation:correl}
\end{eqnarray}

where the function $g$  has a Gaussian  shape, for simplicity.  It  is
defined over a  box  of W pixels centered   on the $j^{th}$   spectral
resolution element.   Its  Full Width at  Half Maximum  (FWHM) $w_g  =
2\sqrt{\frac{\ln 2}{b_g}}$ defines the typical  scale length where the
correlation applies.  The  function is normalized to  a  total flux of
one and is simply written as

\begin{eqnarray}
 g_{j} & = &\frac{1}{G} e^{-b_{g}\cdot j^2}, \\
 & & \nonumber \\
 G     & = &\sum_{j=-W/2}^{W/2} g_{j} \nonumber
\label{equation:gauss}
\end{eqnarray}

The final algorithm we    propose  for the spatial deconvolution    of
spectra therefore involves the minimization of the function

\begin{eqnarray}
\Ccal = \lambda \Ccal_{\chi^2} + \Lcal_1 + \mu \Lcal_2
\label{equation:algospec}
\end{eqnarray}

The Lagrange multipliers $\lambda$ and $\mu$ should  be chosen so that
the deconvolved spectrum matches the  data correctly, in the $\chi^2$,
once it is  re-convolved   with  the  PSF $\Scal(\vec{x})$.  This   is
described for image deconvolution in Courbin et  al.  (1997, 1998) and
Burud et  al. (1998a,b) and in the  following section for the specific
case of spectra.

\section{Simulations}

The algorithm is tested on two different types of simulated data.  The
first   simulation involves blends of   point  sources only, while the
second   one also considers   extended sources, hence illustrating the
capability of the algorithm to unveil faint extended objects hidden by
much   brighter ones.  Both  simulations  include  the effects of slit
misalignment, seeing variation as a function of wavelength, as well as
exaggerated atmospheric refraction.  Gaussian photon noise and readout
noise are also added to the data.  A typical signal-to-noise ratio for
the  data  is  200-300   per    spectral  resolution element.      The
deconvolutions are  performed as in  imaging and the optimal choice of
the different Lagrange multipliers to be used is  guided by the visual
inspection of  the  residual maps  (RM).  The   RM is  the  difference
between the raw data and the deconvolved image re-convolved by the PSF
$\Scal(\vec{x})$, in units  of the  noise.  An accurate  deconvolution
should therefore leave a flat RM with a mean value of 1.

\subsection{Blended point sources}

The two point sources included in the simulations are separated by two
pixels and  are observed  with  a resolution of  4  pixels  FWHM.  The
deconvolved spectra  have  a resolution of  2 pixels  FWHM.   Figure 1
compares the spatial profiles of  the simulated data before and  after
deconvolution. In all simulations, the PSF used is {\sl not assumed to
be    perfectly known}.   It is   determined  from the   spectrum of a
simulated star by applying the PSF construction algorithm described in
Section 4.

\begin{figure}[t] 
\begin{center}  
\leavevmode
\psfig{file=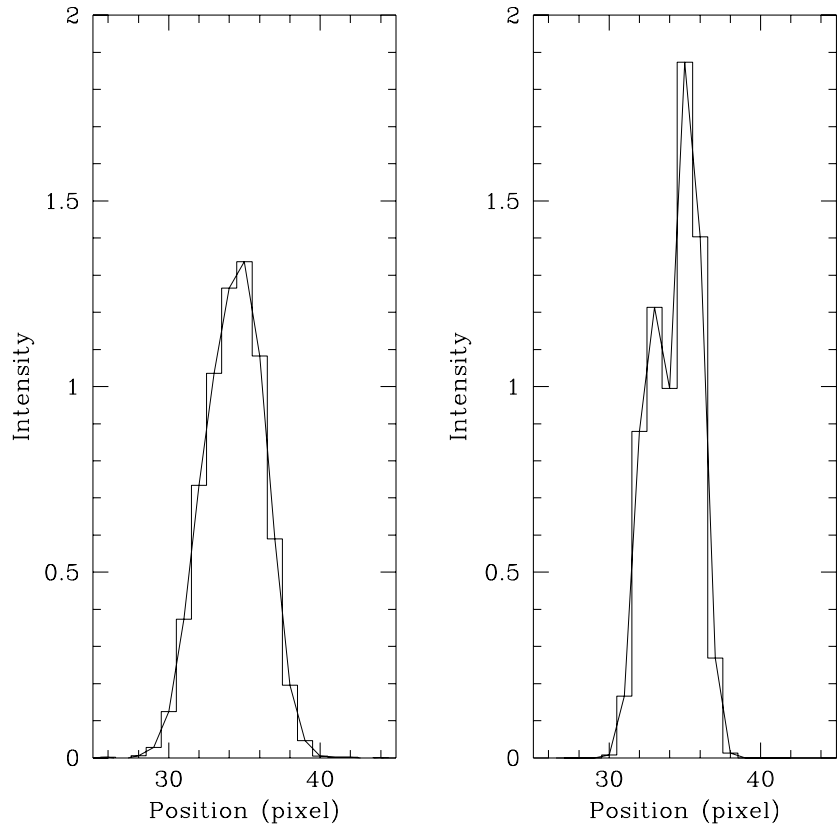,height=12.cm}          
\end{center}
\caption{One-dimensional spatial  profile  of  the  simulated spectrum
before (left) and after  (right) deconvolution. The two  point sources
are perfectly recovered.}
\label{fig:fig1}
\end{figure}

\placefigure{fig:fig1}

We first test the algorithm on a  blend of two very different objects,
with extreme   and  opposite colors.  For  example,   Figures 2 and  3
consider the spectrum  of a quasar contaminated  by that of  a star of
similar or fainter  luminosity (from 0.  to  1.8 mag fainter depending
on  wavelength).  Figure 2  shows the  result of  the deconvolution as
well as the RM which reflects the good  agreement between the data and
the deconvolved spectrum   all along the wavelength range  considered.
Figure 3 displays the 1-D spectra obtained by integrating the light of
the  2-D spectra along the  spatial direction, together  with the flux
ratio between the  deconvolved spectra and the  spectra used to  build
the simulated data (insets).  The latter clearly demonstrates that (1)
the simulated and recovered spectra  agree very well within the  noise
and (2) there is no mutual light contamination between the spectra.

\begin{figure}[t] 
\begin{center}  
\leavevmode
\psfig{file=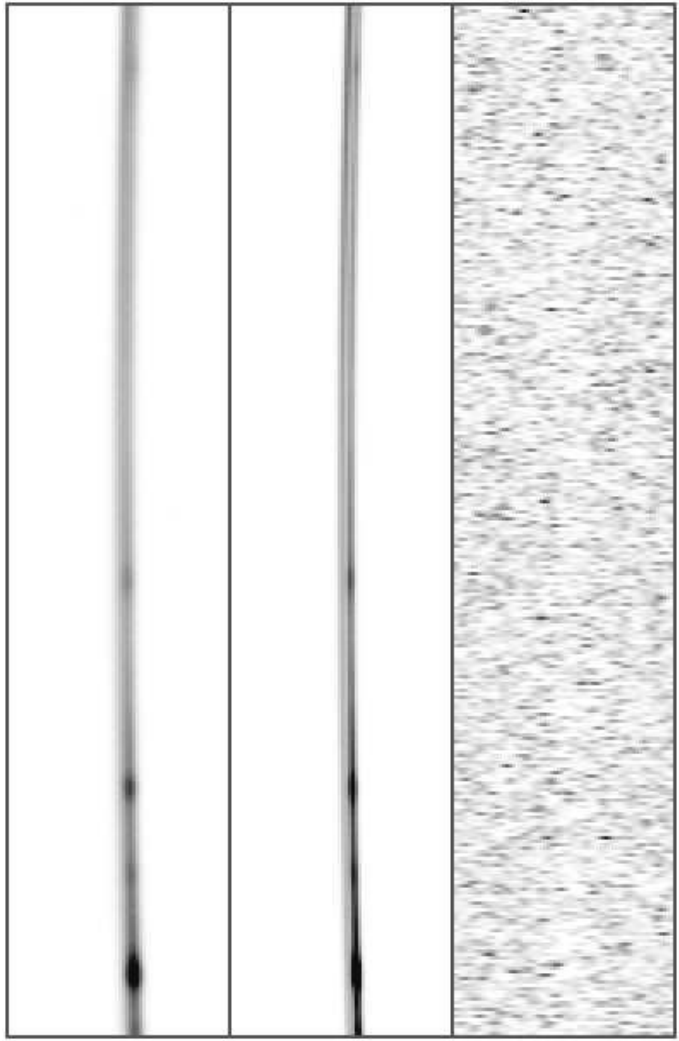,height=14.cm}          
\end{center}
\caption{ Left: Two-dimensional simulated spectrum of a blend of point
sources separated by 2 pixels.  The spatial resolution of the spectrum
is  4   pixels FWHM  and its   curvature  simulates strong atmospheric
refraction.    Middle: 2-D deconvolved   version of  the spectrum. Two
spectra are now -- almost -- resolved. An object with a continuum-only
spectrum can be   seen  on the left    side of the  spectrum  while an
emission   line  object  dominates   the  total  flux    on the  right
side.  Right:  Residual    map   used as  a  quality    check  of  the
deconvolution. It is flat and equal to 1$\sigma$.}
\label{fig:fig2}
\end{figure}

\begin{figure}[t] 
\begin{center}  
\leavevmode
\psfig{file=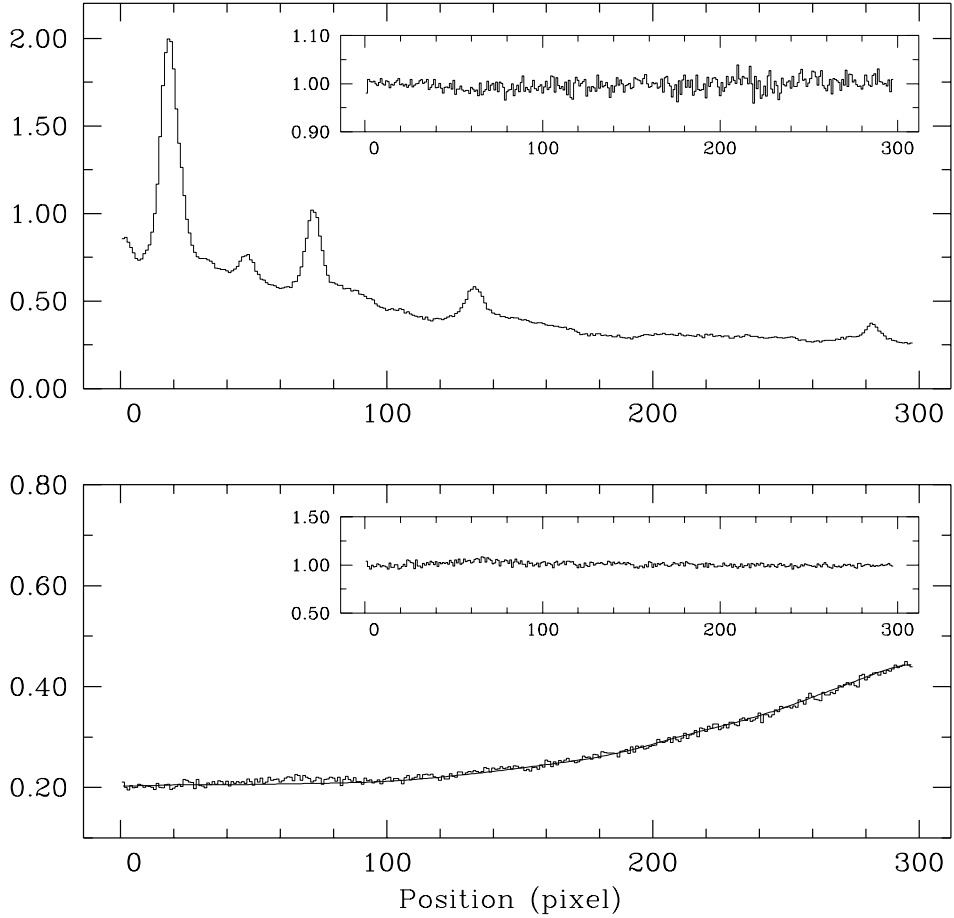,height=14.cm}          
\end{center}
\caption{ Top:  deconvolved 1-D spectrum  of one of the  blended point
source  (a  quasar).  Bottom:  deconvolved 1-D  spectrum of the second
point   source with  only a   continuum.   In both   figures the inset
displays the  flux  ratio between the known   spectrum  used build the
simulated    data and the    deconvolved spectrum.  The  scale on  the
horizontal axis is arbitrary, in pixels.}
\label{fig:fig3}
\end{figure}

\placefigure{fig:fig2}
\placefigure{fig:fig3}

A second test involves a blend of point sources with identical spectra
but different luminosities, e.g., the images  of a lensed quasar.  Our
simulation consist of  two  quasar images with magnification  ratio of
1.8 mag.  As in our first  example, the separation between the sources
is 2  pixels  and the spectra have  a  spatial resolution  of 4 pixels
FWHM.   The 1-D  deconvolved   spectra  are  displayed  in  Figure  4,
confirming  the results obtained  with our first simulation.  Figure 4
also illustrates the effect of  the correlation introduced in equation
\ref{equation:correl} on the position of the  sources as a function of
wavelength.   While  the deconvolved spectra  on  the left panels were
obtained without  introducing  any  correlation ($\mu=0$   in equation
\ref{equation:algospec}), the ones  on the right panels  were obtained
by choosing  the $\mu$  parameter  leading to   the best possible  RM.
Choosing a  too  small $\mu$ multiplier leads   to over-fitting of the
data  and to a ``noisy''  deconvolved  spectrum, while a larger  $\mu$
leads to under-fitting.

\begin{figure}[t] 
\begin{center}  
\leavevmode
\psfig{file=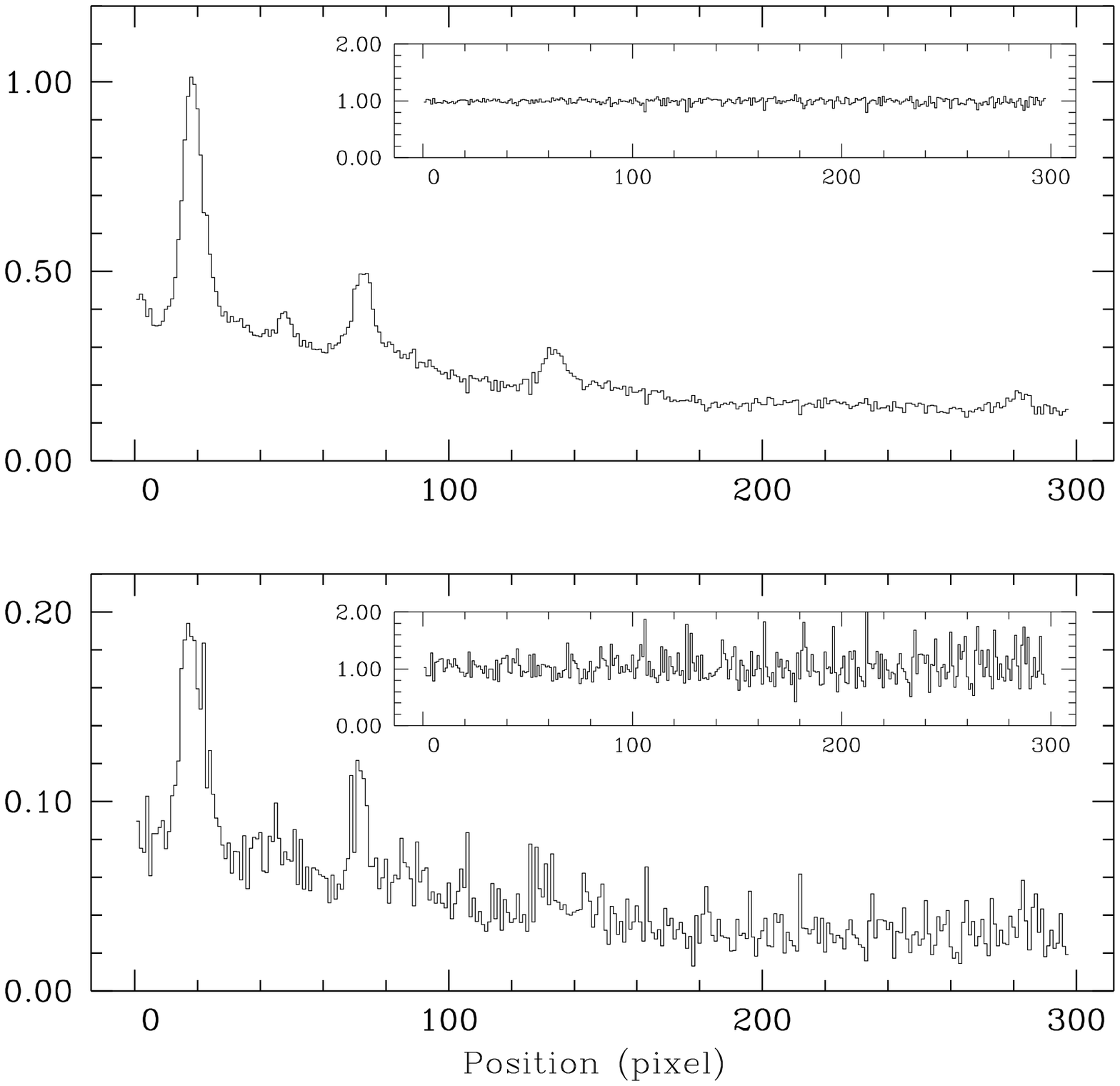,height=8.cm}          
\psfig{file=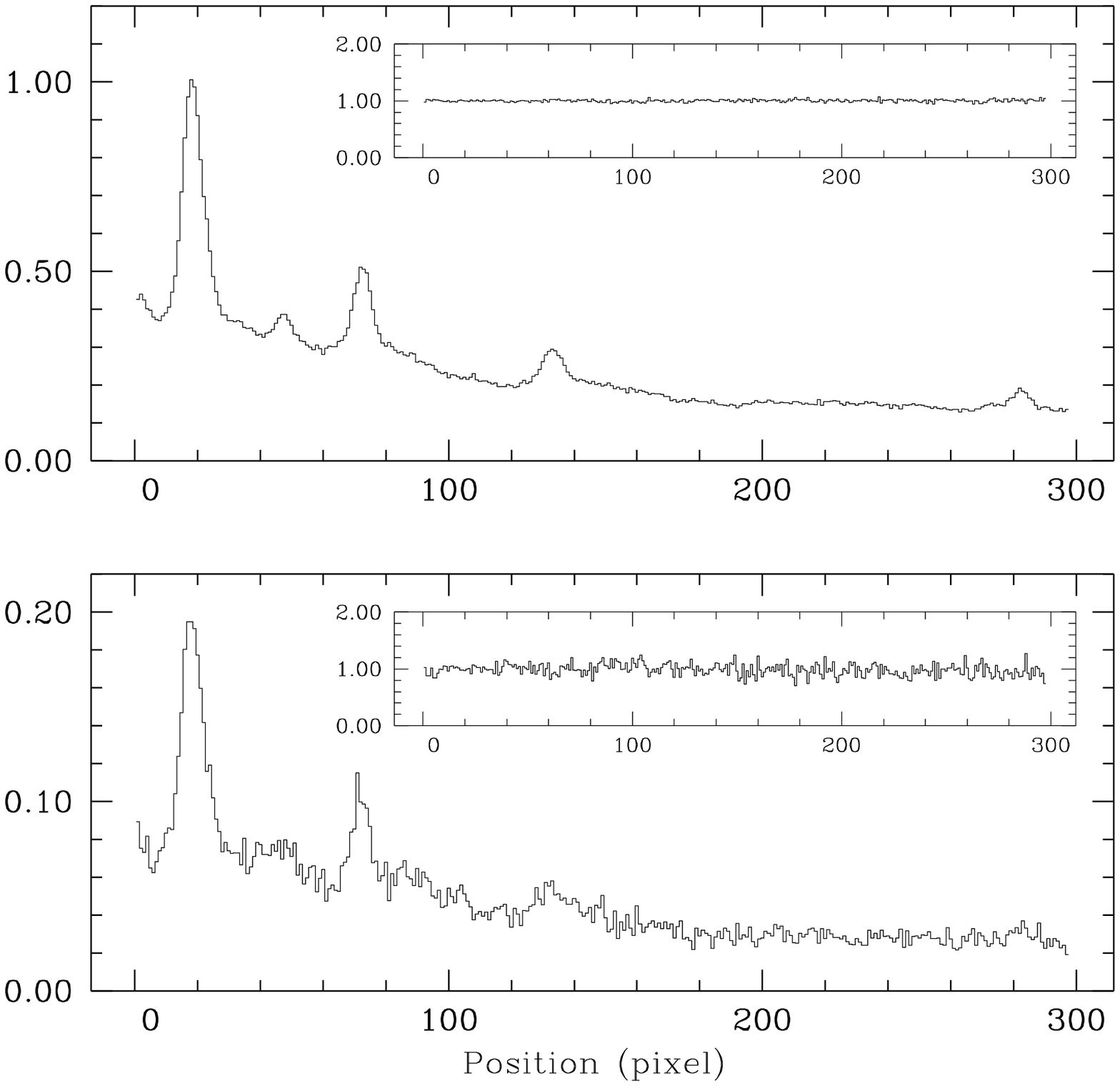,height=8.cm}          
\end{center}
\caption{Left: 1-D  deconvolved QSO spectra, where the  inset is as in
Figure 3.   The deconvolution is performed  with $\mu=0$ (see equation
\ref{equation:algospec}).   Right: As on the  left  panel but with the
correct  choice   for    $\mu$  (see section    3.1).  Note  how   the
signal-to-noise improves with an appropriate choice of the correlation
parameter $\mu$.}
\label{fig:fig4}
\end{figure}

\placefigure{fig:fig4} 

\subsection{Extended sources}

The  simulations presented in  the    previous section show that   the
algorithm   is efficient   in  deconvolving/extracting  the individual
spectra of very blended  point sources  and  that their relative  flux
distribution is not modified by the deconvolution process. We now show
how the algorithm can also be used to extract the spectrum of extended
faint objects hidden by (eventually much brighter) point sources.

\begin{figure}[t] 
\begin{center}  
\leavevmode
\psfig{file=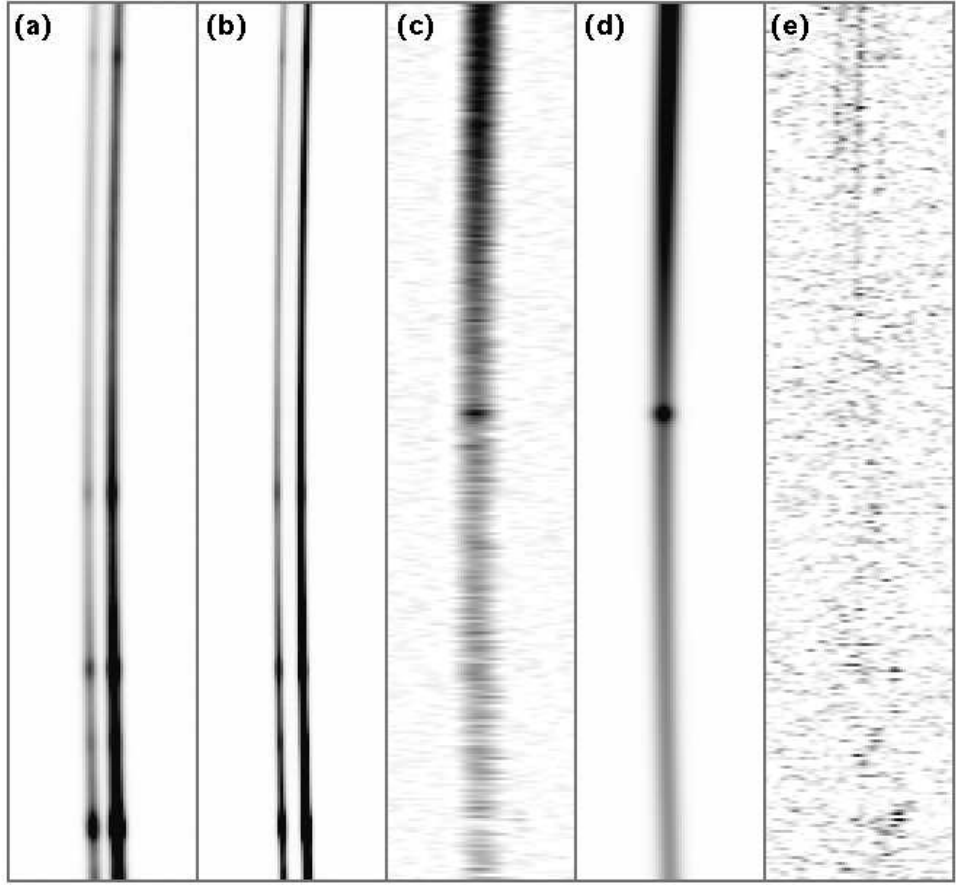,height=12.cm}          
\end{center}
\caption{From left to right: ({\sl a}) Simulated  spectrum of a lensed
quasar  ($\sim 4000-8000$\AA), ({\sl b})  its deconvolution, ({\sl c})
the deconvolved background (here, the lensing galaxy alone), ({\sl d})
the  galaxy  used to  construct  the  simulation,  and ({\sl   e}) the
residual image (data  minus model) in units  of the photon noise.  The
simulation includes the  effects  of seeing variation with  wavelength
and (exaggerated) atmospheric refraction.}
\label{fig:fig5}
\end{figure}

\begin{figure}[t] 
\begin{center}  
\leavevmode
\psfig{file=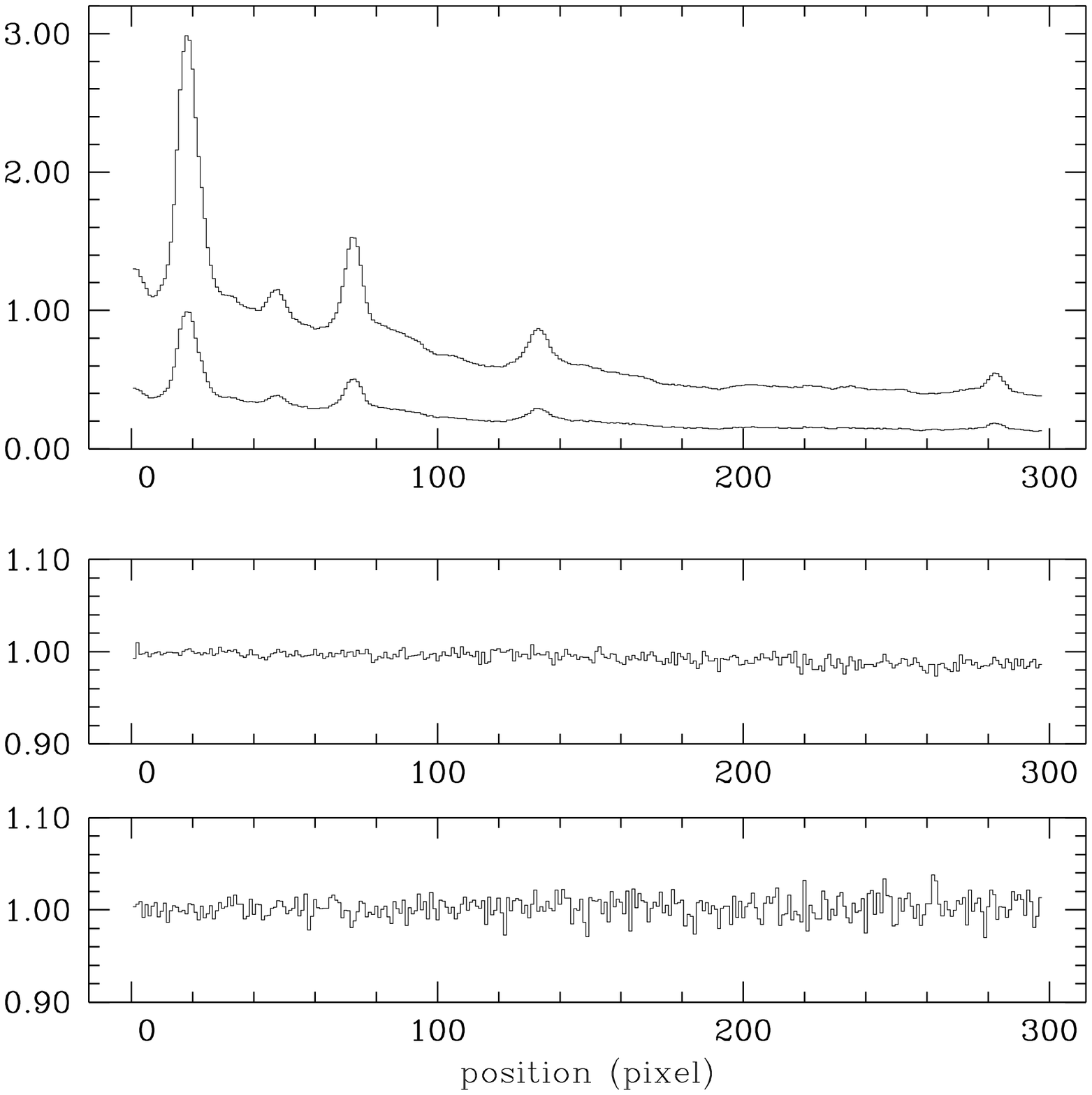,height=8.cm}          
\psfig{file=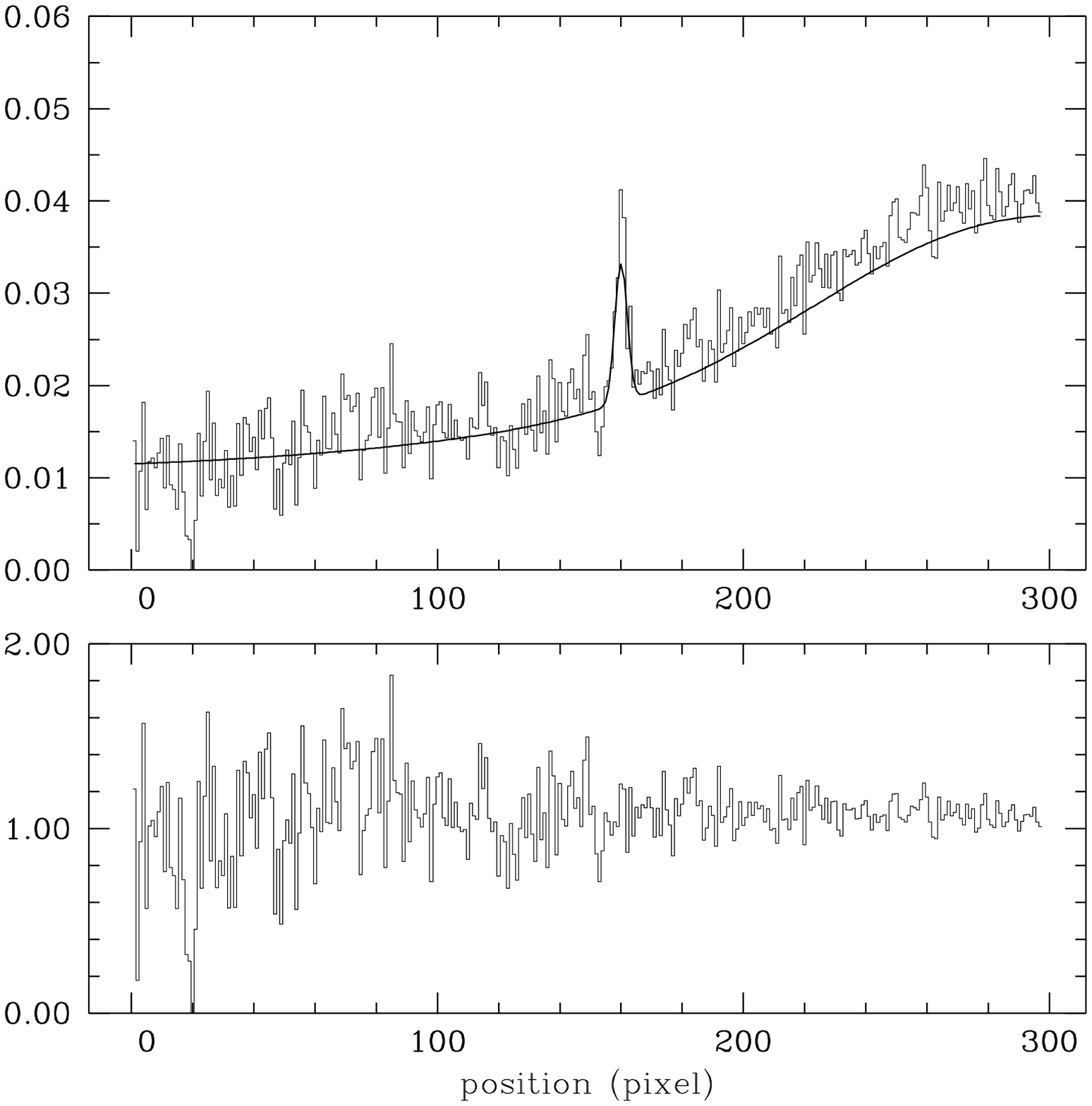,height=8.cm}          
\end{center}
\caption{Left: The top panel shows  the 1-D deconvolved spectra of the
two simulated  QSO images.    In  the middle  and bottom   panels  are
displayed the flux ratio between the two spectra  and the original QSO
spectrum used in  the simulation. Right: the  spectrum  of the lensing
galaxy alone (top) and  its division by  the input spectrum  (bottom).
The position of the emission line is  retrieved with a accuracy of 0.1
pixel.}
\label{fig:fig6}
\end{figure}

\placefigure{fig:fig5}
\placefigure{fig:fig6}

Figure 5  presents the results of  such a simulation.   The spectra of
two quasar images  are generated, with a  separation of 6 pixels.  The
spatial resolution is  4 pixels FWHM  and the signal-to-noise ratio of
the   brightest spectrum is   about  200-300  per spectral  resolution
element,  depending on wavelength.   The  flux ratio  between the  QSO
images  is 3 (1.2  magnitudes).   The spectrum of   the -- extended --
lensing galaxy is also  incorporated  in the  simulated data.  It   is
situated only  2 pixels  away from  the  centroid of the  faintest QSO
image and is about  3  to 5 magnitudes  fainter  than the QSO   images
(depending on wavelength)  and therefore  completely invisible in  the
raw data (see  panel (a) of Figure 5).   Panel (b) shows the result of
the deconvolution, panel (c) displays the  2-D deconvolved spectrum of
the lensing galaxy  alone.  This spectrum is our  best result  among a
number of  deconvolutions  using different smoothness  intensities and
different correlation factors on  the   center of the point    sources
($\lambda$ and $\mu$ in equation \ref{equation:algospec}).  The RM, as
defined at the beginning of this section is displayed in panel (e) and
does   not show   any significant  structure,   as   expected for   an
appropriate choice of $\lambda$ and $\mu$.  Figure 6 confirms the good
results obtained in Figure 5.  The 1-D spectra of  the 2 QSO images as
well as the spectrum of the very faint lensing galaxy are in very good
agreement with  the input spectra,  in spite of  the blending and high
luminosity contrast.  The emission line in the spectrum of the lensing
galaxy is well recovered and its (spectral) position is retrieved with
an accuracy of 0.1 pixel.

\section{Generating the PSF}

Deconvolving spectra requires a good knowledge of the instrumental PSF
all along the wavelength range available.  This condition is fulfilled
as soon as  the  spectrum of  a star or    any other point   source is
recorded  together   with the spectrum    of scientific interest.  The
construction of the PSF is carried out as with the image deconvolution
algorithm, i.e., the  PSF $\Scal(\vec{x})$ is modeled  as the sum of a
Moffat profile and of a numerical image  containing all small residual
differences  between the Moffat and  the observed PSF.  The analytical
one dimensional spatial profile  at  wavelength $j$ is simply  written
as:
 
\begin{eqnarray}
\Mcal_j(\vec{x}) = a_j\, \left[1 + \, b_j\,(\vec{x}-c_j)^{2} \, 
\right]^{-\beta},
\label{equation:moffspec1}
\end{eqnarray}

where $a_j$ is the intensity of the profile,  $b_j$ defines its width,
$c_j$ is    its  center along    the  spatial  direction  and  $\beta$
characterizes the  wings of the  profile.  $\Mcal(\vec{x})$  must have
the   resolution   of  the  PSF   $\Scal(\vec{x})$   needed  for   the
deconvolution process   and is  obtained by   minimizing  the $\chi^2$
between  the observed  PSF  $\Tcal(\vec{x})$   and $\Mcal(\vec{x})\ast
\Rcal(\vec{x})$.  As in the MCS deconvolution, $\Rcal(\vec{x})$ is the
{\sl adopted}  shape of the PSF  after deconvolution. The $\chi^2$ can
be written as follow:

\begin{eqnarray}
\chi^{2}_{\Mcal}      &         =    &    \sum_{j=1}^{M}\sum_{i=1}^{N}
\frac{1}{\sigma_{i,j}^2}            \left[          \Rcal(\vec{x})\ast
\Mcal_{j}(\vec{x}-\vec{c_j})       -      \Tcal_{j}(\vec{x}-\vec{c_j})
\right]_{i}^{2}
\label{equation:psfspec1}
\end{eqnarray}

where the  $i$ and  $j$   indices are respectively running  along  the
spatial and  spectral directions.  As   for the deconvolution,  it  is
highly desirable to use any prior knowledge available on the shape and
position  of the PSF spectrum.  The   center $c_j$ of  the spectrum at
wavelength $j$  is highly correlated to  the  position at neighbouring
wavelength.   The same constraint can  be  applied to  the shape
($b$ and  $\beta$ in equation  \ref{equation:moffspec1}) of the Moffat
profile.   Such  constraints  are  taken   into account by  minimizing
equation    (\ref{equation:psfspec1})   together    with the    three
constraints:

\begin{eqnarray}
\Lcal_{1} & = & \sum_{j=1,}^{M} \left [ 
c_{j} - \frac{1}{G}\sum_{j'=-W/2}^{W/2} g_{j'}c_{j+j'}
\right ]^2  \\
\Lcal_{2} & = & \sum_{j=1,}^{M} \left [ 
b_{j} - \frac{1}{G}\sum_{j'=-W/2}^{W/2} g_{j'}b_{j+j'}
\right ]^2  \\ 
\Lcal_{3} & = & \sum_{j=1,}^{M} \left [ 
\beta_{j} - \frac{1}{G}\sum_{j'=-W/2}^{W/2} g_{j'}\beta_{j+j'}
\right ]^2
\label{equation:correl2}
\end{eqnarray}

where  the function $g$  is  the same  as   in equations (5)  and (6).
Constructing the analytical Moffat component  of the PSF can therefore
be done by minimizing the function:

\begin{eqnarray}
\Ccal_1 = \chi^{2}_{\Mcal} + \mu_1 \Lcal_1 +  \mu_2 \Lcal_2 +  \mu_3 \Lcal_3
\label{equation:moffspec}
\end{eqnarray}

The strengths of the correlations introduced on the shape and position
of  the Moffat profile are modified  by the three Lagrange multipliers
$\mu_1$,  $\mu_2$, $\mu_3$.   While their  choice obviously influences
the quality of the fit, it is not a critical parameter.  Indeed, a PSF
is never a perfect Moffat  profile and a  numerical residual image has
to be  added   to the analytical   component of  the PSF.   Thus,  the
parameters  $\mu_1$, $\mu_2$ and  $\mu_3$ have  to  be chosen  so that
$\Mcal(\vec{x})$ matches at best the  PSF, but an additional numerical
component is {\sl   mandatory}  to  build $\Scal(\vec{x})$  with   the
accuracy required  for the deconvolution  algorithm  to work properly.
This  numerical image $\Fcal(\vec{x})$  must not contain any structure
of spatial frequency    above   the highest frequency   contained   in
$\Rcal(\vec{x})$.  It is therefore constructed by minimizing:

\begin{eqnarray}
\Ccal_2 & = & \sum_{j=1}^{M} \sum_{i=1}^{N} 
\frac{\lambda}{\sigma_{i,j}^2} \left[ \Rcal(\vec{x})\ast \Fcal_{j}(\vec{x}) 
- \Kcal_{j}(\vec{x}) \right]_{i}^{2} \nonumber \\
& + & \sum_{j=1}^{M} \sum_{i=1}^{N} \left[ 
\Fcal_{j}(\vec{x}) - \Rcal(\vec{x})\ast \Fcal_{j}(\vec{x}) \right]_{i}^{2},
\label{equation:specresi}
\end{eqnarray}

where,

\begin{eqnarray}
\Kcal(\vec{x}) & = & \Tcal(\vec{x}) - 
\left[ \Rcal(\vec{x})\ast \Mcal(\vec{x})\right]
\label{equation:diff}
\end{eqnarray}

is the numerical component of the PSF.

The Lagrange parameter $\lambda$ is chosen so that $\Fcal(\vec{x})\ast
\Rcal(\vec{x})$  matches at  best $\Kcal(\vec{x})$ in  the  sense of a
$\chi^2$ and the final PSF  is simply the  sum of $\Mcal(\vec{x})$ and
$\Fcal(\vec{x})$. 

The result of the process is a PSF $\Scal(\vec{x})$ which incorporates
seeing variations  as a function of wavelength  and takes into account
atmospheric refraction.  Using such  a  PSF for the  deconvolution  of
spectra affected by the same atmospheric refraction therefore leads to
a deconvolved  spectrum  corrected   both  for seeing  variation   and
for atmospheric refraction.

One may first think that obtaining the spectrum of a reference star is
a  serious  limitation  to the  technique.   However,   while for some
applications   (e.g.,  multiple QSOs) long   slit  spectroscopy may be
difficult since  a suitable PSF star  well  aligned with the different
objects of  interest might not  be  available, observing with  a Multi
Object Spectrograph (MOS) will in most cases not only allow to observe
the blended objects, but also to obtain  simultaneously the spectra of
one  or more  PSFs.   Observing  several   PSF stars has   the further
advantage of allowing substantial improvement of the spatial sampling.
Higher spatial resolution  can  then be  achieved as  well as   a more
accurate point source/background  separation.  In any  case, either in
long slit  spectroscopy or MOS, particular care  should be paid to the
centering of objects and PSFs on the slit(s).  The slit edges clip the
PSF's wings.  Although  our  deconvolution procedure can  handle 
this, clipping has to be similar in the object spectrum and in the PSF
spectrum.    Observing with wide  slits  will  minimize the effect  of
centering errors and PSF clipping.

\section{Limitations: signal-to-noise and sampling}

To any  algorithm there are  limitations.   The present  one is not an
exception  to  the rule.  Also,   it should be   understood that while
improvement of the data  is aimed at,  the  algorithm can not  extract
non-existing information.

Our   simulations show  that  high  S/N  data are  usually required to
achieve accurate  point-source/background separation,  especially when
dealing with strong  blends.  In the most  extreme case of two objects
exactly superposed, e.g., a  QSO and its  host galaxy, the quality  of
the decomposition also depends  on the physical  size of  the extended
object projected on the plane of the sky, as  compared with the seeing
value.  As  clearly  there is  no way to   separate two  or  more PSFs
located exactly  at the same position,  the main  limitation in such a
case remains the seeing of the observations.

In  addition  to the seeing,   the spatial sampling  of  the data also
influences the results. On 4m class telescopes  pixel sizes tend to be
large in order to beat readout noise and to observe  faint objects.  A
common pixel size is 0.25\arcsec or more, which often leads to poor or
even critical sampling, in the case of  near-IR spectrographs or space
instruments.  As  a consequence,  the gain  in spatial  resolution  is
often limited to less than a factor 2 for data  taken with present day
spectrographs.   However,  the situation   is  improving, as  8m class
telescopes  have   much  smaller pixel    sizes,   of the   order   of
0.1\arcsec. In addition, our algorithm can  make use of an oversampled
PSF (in the  spatial direction) which is  obtained from the spectra of
several PSF stars.   Such observations are possible in  MOS mode.  The
spatial information needed  to restore the  PSF spectrum on a  grid of
pixels smaller than in the original data is then available, as the PSF
spectra are not all centered in the same way,  relative to the central
pixel of each slitlet.  According to the sampling theorem, the gain in
resolution  in only limited by   the sampling {\sl  in the deconvolved
spectrum},  not by the sampling  of the  original data.  This sampling
can in   principle be as small as   the user  wishes,  but we restrict
ourselves to a factor two gain. The number of PSF stars to be observed
simultaneously  to  improve further the  sampling  would be too large.
This still allows  substantial improvement of the  spatial resolution,
even for critically sampled data.  Note finally,  that even if the use
of  an oversampled  PSF  allows one to  deconvolve  critically sampled
spectra, better sampling (FWHM $\sim$ 5-6 pixels) leads, as one should
expect,  to much  more reliable  results,  in  particular  in view  of
accurate background/point source separation.

\section{Conclusions}

We have  described  a new method    for the spatial   deconvolution of
spectra which can be used not only to de-blend point sources, but also
to decompose spectra into a sum of point sources and extended sources.
We  have  shown  from  realistic simulations  that  the relative  flux
distribution in such deconvolved spectra is very well recovered, hence
making it possible to perform accurate spectrophotometric measurements
of very blended objects.   In our simulations,  we resolve and extract
the individual spectra  of sources separated by  only 2  pixels, under
seeing conditions of 4  pixels FWHM.  For modern spectrographs mounted
on 8-10m class telescopes, this  is equivalent to sources separated by
0.2-0.3\arcsec     under   0.4-0.6\arcsec    seeing   conditions.  The
signal-to-noise required for the method to work  efficiently is of the
order of 200,    which is presently  easy   to reach in   a  few hours
integration time, even down to magnitudes of the order of 21-22.

Clearly, the  new extension of  the MCS image  deconvolution algorithm
has a  wide field of  applications (see Courbin  et al. 1999  for more
details on how  to use the algorithm in  practise).  The most original
and  promising  applications  may  consist  in  spectroscopic  studies
involving  extended  objects hidden  by  --  often  brighter --  point
sources.   Our   simulations  show   an  example  of   application  to
gravitationally  lensed quasars, where  the redshift  of a  very faint
lensing galaxy can  be measured, hence making it  possible to estimate
H$_0$ from multiply  imaged QSOs with known time  delays.  A similarly
interesting application will be to  take full advantage of the ability
of the algorithm to decompose spectra, in order to carry out the first
systematic spectroscopic study of  quasar host galaxies.  With current
instrumentation mounted  on 8-10m class  telescopes, sufficiently high
signal-to-noise spectra  can be obtained  for low redshift  quasars in
order to derive precise rotation curves of their host galaxy, provided
the spectrum of the bright QSO nucleus can be removed accurately.  The
present spectra deconvolution algorithm is very well suited for such a
purpose  and  may therefore  allow  significant  progress towards  the
measurement  of  the  mass of  the  central  black  hole in  QSO  host
galaxies.

\begin{center}
{\bf {\large APPENDIX}}
\end{center}

We  assume that the spatial deconvolution  of  spectra simplifies to a
number of independent    deconvolutions of   quasi-monochromatic   1-D
images. This assumption makes sense if, ({\it i}) the PSF is stable in
the spatial direction ({\it   ii}) it varies slowly  with  wavelength,
i.e.,  it does  not   show any  significant   changes in the  spectral
direction on a length scale comparable  to the seeing  disk at a given
wavelength and, ({\it iii}) the PSF is separable. 

Conditions ({\it i}) and   ({\it ii}) are usually   fulfilled provided
PSFs stars can be found close to  the objet to deconvolve and provided
very  low spectral resolution is  not aimed at.  Condition ({\it iii})
might be  more difficult to fulfill  exactly. Using the same notations
as  in the main body of  the paper, but  this time in 2 dimensions, we
note $\Scal(x,y)$ the (narrower) PSF at pixel $(x,y)$, where $x$ is in
the spatial direction and   $y$   is in the spectral   direction.
$\Scal$ is separable if it can be written as

\begin{eqnarray}
\Scal(x,y) & = & \Hcal_{y}(x) \Kcal(y),
\end{eqnarray}

where $\Hcal_{y}(x)$ and $\Kcal(y)$ are two 1-D spatial distributions.
The index $y$ refers to possible variations  of the PSF $\Hcal_{y}(x)$
with wavelength.  In other   words,  $\Scal(x,y)$ is not   necessarily
symmetric about its center but, if elongated, must have its major axis
parallel (or perpendicular) to the slit.  In practice, this means that
the  algorithm may not be  fully   applicable if significant  tracking
errors affect the data.

In the following we will  assume that the instrument  used to take the
data  has a  decent    optical quality, operates  at  relatively  high
spectral resolution  and has a reliable  tracking system.  We can then
write the (noise free) intensity of a data pixel $\Dcal(x,y)$ as

\begin{eqnarray}
\Dcal(x,y) = \Scal_{y}(x,y) \ast \Fcal(x,y)
\end{eqnarray}

where  $\Fcal(x,y)$ is the signal  of scientific interest.  Moderately
high  spectral resolution ensures   us that $\Scal_{y}(x,y)$  does not
vary too fast with  wavelength.  We can  therefore  assume that  it is
constant over   a spatial   area  approximately  equal to  the  seeing
disk. We can now consider only one 1-D spectral resolution element $y$
so that $\Scal_{y}(x,y)$ can  be simplified to $\Scal(x,y)$.  Blurring
by the spectral PSF of the spectrograph $\Wcal(y)$ leads to
 
\begin{eqnarray}
\Dcal(x,y) = \int \, \left[ \Scal \ast \Fcal \right](x,y') 
\Wcal(y-y')\,dy' 
\end{eqnarray}

or more explicitly,

\begin{eqnarray}
\Dcal(x,y)  =  \int \int \int \Scal(x-x",y'-y")\Fcal(x",y") \, dx"
       \, dy" \Wcal(y-y') \, dy' 
\end{eqnarray}

Now, we can write, 

$$\Lcal(x-x",y-y") = \int \, \Scal(x-x",y'-y")\Wcal(y-y') \, dy'$$

to obtain, 

\begin{eqnarray}
\Dcal(x,y) = \Lcal(x,y) \ast \Fcal(x,y) 
\end{eqnarray}

Therefore, if  $\Scal(x,y)$    is separable,   $\Lcal(x,y)$   is  also
separable,  so that  we   can   finally  write $\Dcal(x,y)$ as     the
convolution of a spectrum with a 1-D profile in the spatial direction:

\begin{eqnarray}
\Dcal(x,y) & = &   \Hcal(x) \ast 
\left[ \Kcal(y) \ast \Fcal(x,y) \right]
\end{eqnarray}

with 
\begin{eqnarray}
\Lcal(x,y) = \Hcal(x) \Kcal(y)
\end{eqnarray}

The function $\Kcal(y) \ast \Fcal(x,y)$ is  the spectrum of scientific
interest, and  $\Hcal(x)$  has  here  the  same role as   the  1-D PSF
$\Scal(x)$ used in the algorithm.

\clearpage

\acknowledgments

F.  Courbin    acknowledges   financial support   from  Chilean  grant
FONDECYT/3990024.  M. Kirkove and  S.  Sohy are supported by contracts
ARC~94/99-178 ``Action de   Recherche Concert\'ee de   la Communaut\'e
Fran\c{c}aise     (Belgium)''        and     ``P\^ole     d'Attraction
Interuniversitaire'' P4/05 (SSTC  Belgium). We also thank the European
Southern Observatory for additional support.
 
{}

\clearpage

\clearpage

\clearpage

\clearpage

\clearpage

\clearpage

\begin{thebibliography}{}

\bibitem[Burud et al.   1998a]{bu98a} Burud, I., Courbin,  F., Lidman,
C., et al. 1998a, \apj, 501, L5

\bibitem[Burud et al. 1998b]{bu98b} Burud, I., Stabell, R., Magain, P., 
et al. 1998b, A\&A, 339, 701

\bibitem[Courbin et al. 1997]{cou97} Courbin, F., Magain, P., Keeton, C.R.,
et al. 1997, A\&A, 324, L1

\bibitem[Courbin et al. 1998]{cou98} Courbin, F., Lidman, C., Frye, B., 
et al. 1998, \apj, 499, L119

\bibitem[Courbin et al. 1999]{cou99} Courbin, F., Magain, P., Sohy, S.,
Lidman, C., Meylan, G., 1999, ESO-Messenger, in preparation

\bibitem[Lucy, 1974]{lu74} Lucy, L. 1974, \aj, 79, 745

\bibitem[Magain et al. 1998]{ma98} Magain, P., Courbin, F., \& Sohy, S. 1998, 
\apj, 494, 472

\bibitem[Richardson, 1972]{ri72} Richardson, W.H.J. 1972, 
J. Opt. Soc. Am., 62, 55

\bibitem[Skilling \& Bryan, 1984]{ski84} Skilling, J., \& Bryan, R.K. 1984,
\mnras, 211, 111

\end{thebibliography}
\end{document}